\documentclass[prl,twocolumn, showpacs,superscriptaddress]{revtex4}
\usepackage{graphicx}
\usepackage{amstext}
\usepackage{amsmath}
\usepackage{amssymb}
\usepackage{verbatim}
\usepackage{psfrag}
\usepackage{dcolumn}

\newcommand{\ud}{\mathrm{d}}
\newcommand{\bra}[1]{\langle #1|}
\newcommand{\ket}[1]{| #1\rangle}
\newcommand{\vect}[1]{\boldsymbol{#1}}
\newcommand{\eq}[1]{Eq.~\eqref{#1}}
\newcommand{\fig}[1]{Fig.~\ref{#1}}

\newcommand{\be}{\begin{equation}}
\newcommand{\ee}{\end{equation}}
\newcommand{\ti}[1]{\text{#1}}
\newcommand{\mc}[1]{\mathcal{#1}}
\newcommand{\w}{\omega}
\newcommand{\mean}[1]{\langle #1 \rangle}

\begin{document}

\title{Robust ultrafast currents in molecular wires through Stark shifts }
\author{Ignacio Franco}
%\email[]{ifranco@chem.utoronto.ca}
\affiliation{Chemical Physics Theory
Group, Department of Chemistry, and  Center for Quantum Information and
Quantum Control, University of Toronto, Toronto, Ontario, Canada.}
\author{Moshe Shapiro}
\affiliation{Chemical Physics Department, The Weizmann Institute, 
Rehovot, Israel, and Depts. of Chemistry and Physics, The University of 
British Columbia, Vancouver, B.C., Canada}
\author{Paul Brumer}
%\email[]{pbrumer@chem.utoronto.ca}
\affiliation{Chemical Physics Theory
Group, Department of Chemistry, and  Center for Quantum Information and
Quantum Control, University of Toronto, Toronto, Ontario, Canada.}

\date{\today}

\begin{abstract}
A novel way to induce ultrafast currents in molecular wires
using two incident laser frequencies, $\w$ and $2\w$, 
is demonstrated. The mechanism relies on Stark shifts, 
instead of photon absorption, to transfer population to the excited states and 
exploits the temporal profile of the  field to generate phase controllable 
transport. Calculations in a \emph{trans}-polyacetylene oligomer coupled to metallic 
leads indicate that the mechanism  is highly efficient and  robust to  
ultrafast electronic dephasing processes induced by vibronic couplings. 
\end{abstract}

\pacs{85.65.+h, 33.80.-b, 73.63.-b, 72.10.Di}
\maketitle

Considerable effort has been devoted to studies of the properties of molecular 
wires~\cite{reviews}, as the basic 
building blocks of nanoscale electronics.  The focus is normally 
placed  on the 
transport properties of  metal-molecule-metal junctions  subject to a bias 
voltage. In this regime, the  metallic leads are the main source of electrons 
for  the transport while  the molecular system serves as a transporting medium 
that can be chemically functionalized to  modify the $I$-$V$ characteristics 
of the junction. 

Here we consider an alternative situation in which the junction  is not subject
to a bias voltage and where the molecule serves as the main source of 
transporting electrons. The composite system is taken to be spatially 
symmetric and  net currents   are induced using lasers with  frequency components  $\w$ and $2\w$.  
Such fields  give rise to phase-controllable transport  in symmetric 
systems~\cite{paul, 1vs2, francoprl} even when they have a zero temporal mean.  This 
rectification effect  first appears in the third order response of the system 
to the incident  radiation~\cite{francoprl}. The setup is of interest in 
molecular wires since it can produce ultrafast currents and could lead to the 
development of molecular switches that operate on a femtosecond timescale.

In the usual way that the scenario is applied  the frequencies of the 
$\w+2\w$ field are chosen at or near resonance.  The resulting   multiphoton 
absorption processes serve the double purpose of creating  laser-induced 
rectification and  transferring population to transporting  states. Such 
setup, however, is fragile to decoherence processes since it relies on 
creating  coherent superposition states~\cite{paul}. This aspect becomes particularly  
troublesome when applying the scenario to  molecular nanojunctions as  most 
molecules exhibit strong electron-vibrational 
couplings~\cite{sergei2, reviews}  that  introduce ultrafast coherence loss
(in less than 10 fs~\cite{rossky}) and internal relaxation mechanisms in the electronic dynamics~\cite{thesis}. 

In fact, recent calculations~\cite{thesis} show that vibronic  
couplings can make the laser rectification in molecular wires extremely 
inefficient. In that study we found that less than $4\%$ of the  photoexcited 
electrons  participated in the current, whereas equivalent  but rigid systems 
exhibited 60\% efficiencies.  Hence,  in order to generate sizable currents  
considerable energy from the laser field needs to be dumped into the 
nanojunction compromising  its structural integrity.  Further, employing 
faster or stronger pulses  is not very helpful~\cite{thesis} since they 
introduce other satellite channels in the control  and do not appreciably 
overcome the deleterious effects of the vibronic  couplings.

In this Letter we present an alternative mechanism that  is remarkably robust 
to electron-vibrational couplings, survives in the presence of decoherence and 
thermal effects,  and is able to induce large currents in molecular wires 
with  efficiencies $>90\%$. Instead of promoting electrons to the conduction 
band through absorption of photons, we work far off resonance and employ  
Stark shifts to nonadiabatically couple  the ground and excited electronic 
states. Phase-controllable symmetry breaking is achieved by exploiting the 
difference in the intensity of $\w+2\w$ fields for positive and negative 
amplitudes.

As a system we consider a \emph{trans}-polyacetylene (PA) oligomer connected 
by its ends to macroscopic metallic leads.  The leads are treated as rigid 
semi-infinite tight binding chains while the oligomer is described using the 
well-known Su-Schrieffer-Heeger (SSH) Hamiltonian~\cite{SSH}. The SSH model treats the molecule as a 
tight-binding chain in which the $\pi$ electrons are coupled to distortions in 
the polymer backbone by electron-vibrational interactions. It is  remarkably 
successful in reproducing the electronic structure and dynamics of excitations 
of PA. 

The photoinduced electron-vibrational  dynamics of the nanojunction is 
followed in a mean-field  (Ehrenfest) mixed quantum-classical approximation where transitions between instantaneous eigenstates are allowed~\cite{streitwolf}. The effect of lattice 
fluctuations is incorporated by propagating  an ensemble of quantum-classical 
trajectories with initial conditions selected from the 
ground-state nuclear Wigner distribution function of the chain. 
In this way the dynamics reflects the initial nuclear quantum distribution and 
is subject to the level broadening and internal relaxation mechanism introduced
by the vibronic couplings in the wire~\cite{thesis}. We have found that this 
method describes the trends observed in a recent sub-10 fs experiment performed on PA~\cite{kobayashi}. In addition, we introduce 
projective lead-molecule couplings so that basic aspects of the Fermi sea are 
taken into account without explicitly following the dynamics of the essentially
infinite number of particles in the leads. 

The nanojunction is described as a one-dimensional lattice in which each site 
$n$ corresponds to the position of an atom, and is defined by the Hamiltonian
%------------------------------------------------------------%
\be
\label{eq:tothamcomposite}
H(t) = H_\ti{L} + H_\ti{S-L} + H_\ti{S}(t) + H_\ti{S-R} + H_\ti{R}.
\ee
%------------------------------------------------------------%
Here $H_\ti{L} = -t_\ti{lead}\sum_{n<0, s} 
c_{n+1,s}^\dagger c_{n,s} + \text{H.c.}$ and 
$H_\ti{R}  = -t_\ti{lead}\sum_{n>N, s} 
c_{n+1,s}^\dagger c_{n,s} + \text{H.c.}$ describe the left (L) and right (R) 
lead with  hopping parameter  $t_\text{lead}$. The operator $c_{n,s}^\dagger$ 
($c_{n,s}$) creates (annihilates) a fermion in site $n$ and spin $s$, and H.c. 
denotes the Hermitian conjugate.  The SSH oligomer with Hamiltonian 
$H_\ti{S}(t) = H_\ti{el}  + H_\ti{latt}$  is situated between sites 
$n=1, \cdots, N$ and is coupled to  an electric field $E(t)$ in the dipole 
approximation. Specifically, the electronic part of $H_\ti{S}$ is
\be
\begin{split}
H_\text{el} = \sum_{n=1,s}^{N-1} [-t_0+\alpha(u_{n+1}-u_n)] 
(c_{n+1,s}^\dagger c_{n,s} \\
+ c_{n,s}^\dagger c_{n+1,s}) + |e| \sum_{n=1,s}^{N} x_{n} c_{n,s}^\dagger 
c_{n,s} E(t),
\end{split}
\ee
 where  $t_0$ is the hopping integral for zero displacement, $u_n$ is the 
monomer displacement of site $n$ and $\alpha$ describes the electron-ion 
coupling between neighboring sites. In turn, $x_n=(na+ u_n)$ is the position 
operator for site $n$,  $a$ the lattice constant, and  $-|e|$  the unit 
electronic charge.   The lattice is described by 
%------------------------------------------------------------%
\begin{equation}
H_{\textrm{latt}} = \frac{M}{2}\sum_{n=1}^{N} \dot{u}_n^2 + 
\frac{K}{2}\sum_{n=1}^{N-1} \left(u_{n+1} - u_{n}\right)^{2} 
-|e|\sum_{n=1}^{N} x_n E(t),
\end{equation}
%------------------------------------------------------------%
with force constant $K$ and (CH) group mass $M$. 

We consider the case where  no bias voltage is applied across the bridge and 
focus on  the short time dynamics of the system. In this regime,  the basic function  
of the leads is to absorb electrons from the molecule with energy higher than 
the Fermi energy $\epsilon_\ti{F}$, taken to be the zero-reference energy, and 
to block them otherwise. We focus on  electronic transport and model this effective 
coupling using projection operators.  Specifically,  the  lead-molecule 
coupling 
%------------------------------------------------------------%
\be
\label{eq:projectedcoupling}
\begin{split}
H_\ti{S-L}(t)    & = -t_\ti{coup} \sum_{n\in \ti{S}, s} \mc{P}^\ti{S}_{1,n}(t) 
c_{0,s}^\dagger c_{n,s} +  \text{H.c.}; \\
H_\ti{S-R}(t)     &= -t_\ti{coup} \sum_{n\in \ti{S}, s}  \mc{P}^\ti{S}_{N,n}(t)
c_{N+1,s}^\dagger c_{n,s} + \text{H.c.},
\end{split}
\ee
%------------------------------------------------------------%
is a  restricted tight-binding interaction with coupling constant  
$t_\text{coup}$  in which  only electrons with energy 
$\epsilon>\epsilon_\ti{F}$ are  deposited in the contacts. Here   
$\mc{P}^\ti{S}(t)=\sum_{\epsilon_\gamma> \epsilon_\ti{F}} 
\ket{\gamma_\ti{S}(t)}\bra{\gamma_\ti{S}(t)}$  projects into the instantaneous 
molecular $\pi^\star$ light-dressed eigenorbitals defined  by the eigenvalue problem $H_\ti{el}(t)\ket{\gamma_\ti{S}(t)} = \epsilon_\gamma \ket{\gamma_\ti{S}(t)}$, and  $\mc{P}^\ti{S}_{n,m} = \bra{n}\mc{P}^\ti{S}\ket{m}$ with $\ket{n}=c_{n,s}^\dagger\ket{0}$ where $\ket{0}$ is the vacuum state.  Note that since the SSH model is electron-hole symmetric a subsequent inclusion of hole transport is just expected to double the resulting current.

In the mean-field approximation~\cite{streitwolf} the nuclei are described by 
classical trajectories determined by  
\begin{multline}
\label{eq:class}
M\ddot{u}_{n}  = - \langle \Psi(t)| \frac{\partial H(t)}{\partial u_n}
|\Psi(t)\rangle =
- K\left(2u_n-u_{n+1}-u_{n-1}\right) \\
 + 2\alpha\textrm{Re}\left\{ \rho_{n, n+1} 
-   \rho_{n, n-1} \right\} 
 - |e|E(t) \left( \rho_{n,n} -1\right).
\end{multline}
Here 
$\rho_{n,m} = \sum_{\epsilon, s}     f(\epsilon, s) 
\langle m \ket{\epsilon(t)}\bra{\epsilon (t)}n\rangle$ is the  electronic 
reduced density matrix, and  
$f(\epsilon, s)$   is the time-independent initial distribution function 
(that takes values 0 or 1 depending on the occupation of each level  with 
energy $\epsilon$ and spin $s$). We take the chain to be clamped, so that 
$u_1(t)=u_N(t)=0$ at all times. In turn, the orbitals 
$\ket{\epsilon(t)}$ that form the $\mc{N}$-electron wavefunction  
$\ket{\Psi(t)}$ satisfy the time-dependent  Schr\"odinger equation
%------------------------------------------------------------%
\begin{equation*}
\label{eq:seorbital}
i\hbar \frac{\ud}{\ud t}
\begin{bmatrix} 
\ket{\epsilon_{\ti{L}}(t)} \\ 
\ket{\epsilon_\ti{S}(t)} \\ 
\ket{\epsilon_\ti{R}(t)}
\end{bmatrix}
= 
\begin{bmatrix}
H_{\ti{L}} & H_\ti{S-L}(t) & 0 \\
H_\ti{S-L}(t) & H_\ti{el}(t) & H_\ti{S-R}(t) \\
0 & H_\ti{S-R}(t) & H_\ti{R} 
\end{bmatrix}
\begin{bmatrix} 
\ket{\epsilon_{\ti{L}}(t)} \\ 
\ket{\epsilon_\ti{S}(t)} \\ 
\ket{\epsilon_\ti{R}(t)}
\end{bmatrix},
\end{equation*}
%------------------------------------------------------------%
where   $\ket{\epsilon_{\ti{P}}} =  \sum_{n\in \ti{P}} 
\bra{n}\epsilon\rangle \ket{n}$ describes  the part of the orbital in region 
P=L, R or S. We take the leads and molecule to be initially detached and,  
since the projective coupling [\eq{eq:projectedcoupling}] already incorporates 
the Fermi blockade imposed by the leads, only consider orbitals that are 
initially  in the molecular region, so that  $\ket{\epsilon_\beta(0)}=0$ with 
$\beta=$L or R.  

As the composite system is spatially unbounded, it is necessary to obtain an 
equation of motion for the system part of the orbitals 
$\ket{\epsilon_\ti{S}}$ where we do not  need  to explicitly follow its 
behavior  in the lead regions. This can be accomplished by applying 
Laplace transform techniques to the evolution equation for the lead components,
to obtain:  $\ket{\epsilon_\beta(t)}  =   \frac{1}{i\hbar} 
\int_0^t  U^\beta(t-\tau) H_{\ti{S-}\beta}(\tau) 
\ket{\epsilon_\ti{S}(\tau)}\ud\tau$. Here the evolution operator for the 
isolated leads   $U^\beta(t)=\exp(-iH_\beta t/\hbar)$ acts as  the memory 
kernel of the convolution integral. Its matrix elements in site representation
can be determined in closed form.  For this, note that $H_\beta$ is 
diagonalized by the  basis transformation  
$c_{n,s} = \left(\frac{2}{\mc{M}}\right)^{1/2} 
\sum_{k=1}^{\mc{M}} \sin\left[(n-n_{\beta})k \pi/\mc{M}\right] c_{k,s}$, 
where $\mc{M}$ is the number of lead sites,   $n_\ti{L}=1$  and $n_\ti{R}=N$. 
It then follows that $U_{n,m}^\beta(t)  =  i^{n-m}  J_{n-m}
\left(\frac{2 t_\ti{lead}}{\hbar} t\right) 
- i^{n+m-2n_\beta}  J_{n+m-2n_\beta}\left(\frac{2 t_\ti{lead}}{\hbar} t\right)$
for large $\mc{M}$,  where $J_{n}(z)= \frac{i^{-n}}{\pi}
\int_{0}^{\pi} e^{i z \cos{\theta}} \cos(n \theta)\, \ud\theta$ is a Bessel 
function of the first kind of order $n$.  Hence,
%------------------------------------------------------------%
\begin{multline}
\label{eq:nmseprojected}
i\hbar \frac{\partial }{\partial t} \bra{n}\epsilon_\ti{S}(t)\rangle
  = 
\bra{n} H_\ti{el}(t)\ket{\epsilon_\ti{S}(t)}  \\ 
+ \frac{t_\ti{coup}^2}{i\hbar} \sum_{m=1}^{N} \int_{0}^{t}  \, 
\mc{K}(t-\tau) 
\Gamma_{n,m}(t,\tau)\bra{m}\epsilon_\ti{S}(\tau)\rangle  \ud \tau,
\end{multline}
%------------------------------------------------------------% 
where  $\Gamma_{n,m}(t,\tau)= \mc{P}_{n,1}^{\ti{S}}(t) 
\mc{P}_{1,m}^{\ti{S}}(\tau)+ \mc{P}_{n,N}^{\ti{S}}(t) 
\mc{P}_{N,m}^{\ti{S}}(\tau)$.  Equation~\eqref{eq:nmseprojected}  is an 
effective non-Markovian Schr\"odinger equation for the system part of the 
orbitals.  The first term corresponds to the single-particle Schr\"odinger 
equation for the isolated system. The convolution integral  with memory 
kernel  $\mc{K}(t)=  U_{0,0}^\ti{L}(t)= U_{N+1, N+1}^{R}(t)= 
{2 J_{1}\left(\frac{2 t_\ti{lead}}{\hbar} t\right)}/
{\frac{2 t_\ti{lead}}{\hbar} t}$ describes the transfer of population from 
the system into the leads.

We now invoke the wide bandwidth approximation, where  
$\hbar/2t_\text{lead}$ becomes  the fastest  time scale in the problem. In this
regime,  \eq{eq:nmseprojected} reduces  to its Markovian limit
%------------------------------------------------------------%
\be
\label{eq:finalorbs}
i\hbar \frac{\partial}{\partial t}\bra{n}\epsilon_\ti{S}\rangle = 
\sum_{m=1}^{N}\left[\bra{n} H_\ti{el}(t)\ket{m}  
- i \frac{t_\ti{coup}^2}{t_\ti{lead}} \Gamma_{n,m}'(t)\right]
\bra{m} \epsilon_\ti{S}\rangle,
\ee
%------------------------------------------------------------%
and the leads are effectively mapped into a negative imaginary (absorbing) 
potential.    The projection operators in the coupling, contained within 
$\Gamma_{n,m}'(t)=  \Gamma_{n,m}(t, t)$,  ensure that only those electrons 
with sufficient energy get absorbed with proper conservation of the 
antisymmetry principle. The equation   is valid for times 
$t\gg \hbar/t_\ti{lead}$, but  it can be used  for all times by slowly turning 
on  the lead-molecule interaction; a strategy that we adopt.  Note that the 
field influences the dynamics directly by inducing transitions among states, 
and indirectly by modifying the wire-lead couplings. Equations~\eqref{eq:class}
and~\eqref{eq:finalorbs}  constitute a closed set of $N(N+2)$ first-order 
differential equations that are integrated using a Runge-Kutta method of 
order eight. The projective term in the dynamics $\Gamma_{n,m}'(t)$ is updated 
at every time step by diagonalizing  $H_\ti{el}(t)$.   Throughout we use the standard SSH parameters~\cite{SSH}: 
$\alpha = 4.1$ eV/\AA, $K=21$ eV/\AA$^2$, $t_0=2.5$ eV,  
$M=1349.14$ eV fs$^2$/\AA$^2$ and $a=1.22$ \AA.  In turn, we take the molecule 
and leads to be weakly coupled with $t_\ti{coup}^2/t_\ti{lead}=0.1$ eV.  

Electronic dephasing due to vibronic coupling  is incorporated by integrating 
the equations for  an ensemble of 1000 initial conditions. The initial 
conditions are generated using the following strategy~\cite{thesis}: the starting  optimal geometry is obtained  by minimizing the total ground-state energy of the chain. Then,  a normal mode analysis around this geometry is performed, yielding the ground state nuclear 
Wigner phase space distribution function 
$\rho_\ti{W}(\vect{u}, \vect{p})$ in the harmonic approximation, where 
$\vect{u}= (u_1, \cdots, u_N)$ and $\vect{p}=M(\dot{u}_1, \cdots, \dot{u}_N)$. 
By importance sampling this distribution  an ensemble of lattice initial 
conditions $\{ \vect{u}^i(0), \vect{p}^i(0)\}$ is generated. The associated 
initial values for the orbitals $\{\ket{\epsilon^i(0)}\}$ are obtained by 
diagonalizing $H_\ti{el}$ for each initial geometry of the lattice 
$\{\vect{u}^i(0)\}$.  Each member $i$ of the ensemble defines a 
quantum-classical trajectory  $( \vect{u}^i(0), \vect{p}^i(0), 
\ket{\Psi^i(0)})\to( \vect{u}^i(t), \vect{p}^i(t), \ket{\Psi^i(t)}) $ and the 
set is used to obtain ensemble averages.  We note that the resulting initial state  is  basically stationary under field-free evolution.

The current entering into lead $\beta$ is defined by 
$j_\beta =  - |e| \partial p_\beta/\partial t$,  where  
$p_\beta= \sum_{n\in \beta} \rho_{n,n}$ is the number of electrons in lead 
$\beta$. Using the same set of techniques employed to arrive at 
\eq{eq:finalorbs}, one can obtain an equation for $j_\beta(t)$ that solely 
depends on molecular properties:
$j_\beta(t)  = -\frac{2|e|}{\hbar}\frac{t_\ti{coup}^2}{ t_\ti{lead} } 
\sum_{m,n}
\text{Re}\{\mc{P}_{n_\beta,m}^{\ti{S}}(t) \mc{P}_{n,n_\beta}^{\ti{S}}(t) 
\rho_{n,m}(t)  \}$.
Any rectification generated by $\w+2\w$ pulses  manifests as 
$j_\ti{L}-j_\ti{R}\ne 0$.

%------------------------------------------------------------------%
\begin{figure}[t]
\centering
\psfrag{t (fs)}[][]{\scriptsize{$t$ (fs)}}
\psfrag{E(t)}[][]{\scriptsize{$E(t)$ (V/\AA)}}
\psfrag{jbeta}[][]{\scriptsize{$j_\beta/|e|$ (1/fs)}}
\psfrag{jl}[l][r]{\scriptsize{$j_\ti{L}$}}
\psfrag{jr}[l][r]{\scriptsize{$j_\ti{R}$}}
\psfrag{energy}[][]{\scriptsize{$\mean{\epsilon_\gamma}$ (eV)}}
\psfrag{A}[][]{\scriptsize{(a)}}
\psfrag{B}[][]{\scriptsize{(b)}}
\psfrag{C}[][]{\scriptsize{(c)}}
\includegraphics[width=0.5\textwidth, angle=0]{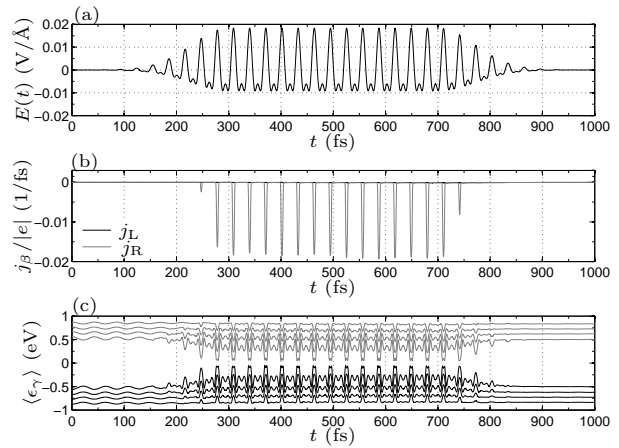}
\caption{Time dependence of (a) the electric field; (b) the current entering 
the left and right leads and; (c) the instantaneous field-dressed orbital energies for states near the energy gap for a 100-site flexible wire under the influence of the field in
\eq{eq:nafield2}  with $\phi_{2\w}-2\phi_\w=0$. Note the bursts of charge 
deposited in the right lead  when the field bridges the energy gap. Here  
$j_\ti{L}$ is so small that is barely visible.}
     \label{fig:flexnotgaussian_0}
\end{figure}
%------------------------------------------------------------------%

We study neutral 100-site chains initially in the ground electron-vibrational 
state. The geometry of the chain consists of a perfect alternation of double 
and single bonds. The electronic structure is composed of 50 doubly occupied 
valence $\pi$ orbitals and 50 empty $\pi^\star$ states,  separated by an 
energy gap of $2\Delta=1.3$ eV. We follow the electron-vibrational  dynamics of
the wire under the influence of an $\w+2\w$ field of the form
%------------------------------------------------------------%
\be
\label{eq:nafield2}
E(t) = \epsilon_{\w}\cos(\w t+\phi_{\w}) +\epsilon_{2\w}\cos(2\w t+\phi_{2\w}).
\ee
%------------------------------------------------------------%
The field is smoothly turned on and off in  $100$ fs and has constant 
amplitude for $400$ fs.   The frequency $\hbar\w= 0.2\Delta$  is chosen far 
off-resonance from the system's interband transition frequencies so that 
Stark shifts, and not photon absorption, dominate the dynamics. The field 
amplitude used is $\epsilon_{2\w} = 6.1\times 10^{-3}$ V \AA$^{-1}$ with 
$\epsilon_{\w} = 2\epsilon_{2\w}$,  which corresponds to an intensity 
$I_{2\w} \sim 5\times 10^{8}$ W cm$^{-2}$.   

Figure~\ref{fig:flexnotgaussian_0} shows the field, the  currents and  
electronic structure of the chain, averaged over all trajectories, when the 
relative phase of the pulse is  $\phi_{2\w}-2\phi_\w=0$. The single-particle 
spectrum displays considerable Stark shifts and the field effectively reduces 
the energy gap of the oligomer, causing frequent crossings between the ground 
and excited electronic states in individual trajectories. Since the wire's 
ground state is non-degenerate and of definite parity, the lowest order 
contributions to the Stark effect is quadratic in the field. Hence, when 
$|E(t)|$ is maximum the energy gap acquires its minimum value.  At the crossing
times  population is transfered from the valence to the conduction band   and  
bursts of charge are deposited in the leads.

Note that for $\phi_{2\w}-2\phi_\w=0$ almost all excited electrons are 
deposited in the right contact only. Symmetry breaking  arises due to the 
difference in the maximum $|E(t)|$ for positive and negative field amplitudes. 
Even when $E(t)$ has a zero temporal mean,  it consists of narrow peaks with 
large $|E(t)|$ for positive amplitudes, and shallow and broad features when 
$E(t)$ is negative (\fig{fig:flexnotgaussian_0}a).  For this reason, the Stark 
effect is only sufficiently strong to close the energy gap  when the field has 
a positive amplitude.  Thus, transfer of population to the conduction band and 
absorption of electrons by the leads always occurs when the laser  is pointing 
at a particular and the same direction,  in this way inducing directed 
transport in the system. 

%------------------------------------------------------------------%
\begin{figure}[t]
\centering
\psfrag{phi/pi}[][]{\scriptsize{$(\phi_{2\w}-2\phi_\w)/\pi$}}
\psfrag{ql-qr}[][]{\scriptsize{$(q_\ti{L}-q_\ti{R})/|e|$}}
\psfrag{eta}[][]{\scriptsize{$\eta$}}
\psfrag{A}[][u]{\scriptsize{(a)}}
\psfrag{B}[][u]{\scriptsize{(b)}}
\includegraphics[width=0.5\textwidth]{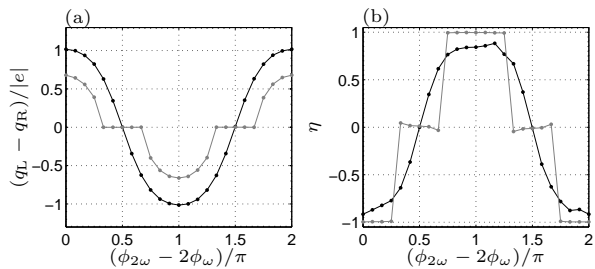}
\caption{Phase dependence of (a) the net rectification $q_\ti{L}-q_\ti{R}$, 
where $q_\beta=\int_{0}^{t_f} j_\beta(t)\ud t$  and; (b) efficiency 
$\eta= (q_\ti{L}-q_\ti{R})/(q_\ti{L}+q_\ti{R})$ of the process when 
field~\eqref{eq:nafield2} is applied to flexible (black dots) and rigid (gray 
dots) 100-site wires. Note the assistance of  phonons in the rectification. }
     \label{fig:notgaussian_phase_dependence}
\end{figure}
%------------------------------------------------------------------%

The phenomenon depends intimately on the relative phase. For instance, for the 
case of $\phi_{2\w}-2\phi_\w=\pi/2$ (not shown) the electric field exhibits 
equal intensity for positive and negative amplitudes. The energy gap is closed
every half period, and since the field changes sign between consecutive 
interband couplings, no net current is produced.

Figure~\ref{fig:notgaussian_phase_dependence} shows the  net difference in 
charge deposited in the left and right leads after the pulse is over  for 
different laser phases, as well as the efficiency of the process. For 
comparison purposes the plot also includes results obtained in an equivalent 
but rigid system. The system is made rigid by increasing the mass of the (CH) 
groups by $10^6$. We first  note that the mechanism is robust to decoherence 
effects due to coupling to the vibrational degrees of freedom as well as  
satellite contributions due to parasite multiphoton absorption.  In fact, 
90\% of the excited electrons can participate of the net current.  Further,  
the sign and magnitude of the effect   can be manipulated  by varying the 
laser phases. By changing the relative phase by  $\pi$ the magnitude of the 
effect stays the same but the direction of the rectification is reversed.

In the flexible wire the rectification exhibits an almost sinusoidal dependence
on $\phi_{2\w}-2\phi_\w$. By contrast, in rigid wires for  certain range of  
phases no currents are induced since the maximum field amplitude is not large 
enough to  couple valence and conduction bands. Hence, in this range the 
currents observed in the flexible wire are phonon-assisted, i.e. the level 
broadening introduced by the vibrations permit the nonadiabatic coupling. The 
currents generated in the flexible case are always larger than in the rigid 
example. However, in rigid wires the mechanism can exhibit perfect 
efficiencies.

In conclusion, we have demonstrated  a way  to efficiently induce 
phase-controllable transport along molecular wires on a femtosecond timescale  
using $\w+2\w$ fields. The effect employs Stark shifts to transfer population 
to transporting states, and exploits the difference in the  instantaneous field
intensity  for positive and negative amplitudes to generate rectification. It 
is robust  to electron-vibrational couplings and other decoherence mechanisms, 
and is best suited for long oligomers.    

%\bibliography{prl}

\end{document}